\documentclass[aps,prl,twocolumn,superscriptaddress]{revtex4-2}

\usepackage{amsmath}
\usepackage{amssymb}
\usepackage{graphicx}
\usepackage{epstopdf}
\usepackage{dcolumn}
\usepackage{bm}
\usepackage{xcolor}
\usepackage[marginal]{footmisc}
\usepackage{flushend}
\usepackage{hyperref}

\hypersetup{hidelinks}

\usepackage{amsmath}
\usepackage{amssymb}
\usepackage{graphicx}
\usepackage{epsfig}
\usepackage{dcolumn}
\usepackage{bm}
\usepackage{xcolor}
\usepackage{hyperref}
\usepackage{csquotes}
\usepackage{balance}

\hypersetup{
    colorlinks=true,
    linkcolor=blue,
    citecolor=blue,
    urlcolor=blue
}

\newcommand{\Br}{{\bf r}}
\begin{document}

\title{Experimental Observation of Ghost Image Revivals via Structured Coherence}

\author{Weining Wang}
\affiliation{Shandong Provincial Key Laboratory of Light Field Manipulation Physics and Applications \& School of Physics and Optoelectronics, Shandong Normal University, Jinan 250358, China}
\affiliation{Collaborative Innovation Center of Light Manipulations and Applications, Shandong Normal University, Jinan, 250358, China}

\author{Yaning Zhou}
\affiliation{Shandong Provincial Key Laboratory of Light Field Manipulation Physics and Applications \& School of Physics and Optoelectronics, Shandong Normal University, Jinan 250358, China}
\affiliation{Collaborative Innovation Center of Light Manipulations and Applications, Shandong Normal University, Jinan, 250358, China}

\author{Sergey A. Ponomarenko}
\email{serpo@dal.ca}
\affiliation{Department of Physics and Atmospheric Science, Dalhousie University, Halifax B3H 4R2, Canada}
\affiliation{Department of Electrical and Computer Engineering, Dalhousie University, Halifax B3J 2X4, Canada}

\author{Yangjian Cai}
\email{yangjiancai@sdnu.edu.cn}
\affiliation{Shandong Provincial Key Laboratory of Light Field Manipulation Physics and Applications \& School of Physics and Optoelectronics, Shandong Normal University, Jinan 250358, China}
\affiliation{Collaborative Innovation Center of Light Manipulations and Applications, Shandong Normal University, Jinan, 250358, China}

\author{Xin Liu}
\email{xinliu@sdnu.edu.cn}
\affiliation{Shandong Provincial Key Laboratory of Light Field Manipulation Physics and Applications \& School of Physics and Optoelectronics, Shandong Normal University, Jinan 250358, China}
\affiliation{Collaborative Innovation Center of Light Manipulations and Applications, Shandong Normal University, Jinan, 250358, China}

\date{\today}

\begin{abstract}
Ghost imaging retrieves an object’s image from intensity correlations between two light beams, neither of which independently carries information about the object. However, conventional ghost imaging critically relies on precise object positioning and conjugate matching between the two arms, causing the image to disappear when these conditions are violated and the object is accessible from one plane only. In this Letter, we break this fundamental limitation by reporting the first experimental observation of revivals of ghost images of arbitrary objects via engineering the longitudinal intensity autocorrelation of a structured thermal light field into a coherence comb. With the reference arm fixed, translating the object produces periodic revivals of the ghost image whenever the object position matches a comb-tooth position, reminiscent of the Talbot effect. This capability enhances the flexibility of correlation imaging, enabling robust tomographic imaging of moving and non-periodic complex objects.
\end{abstract}

\maketitle

Ghost imaging introduces a counterintuitive paradigm in which the image of an object can be reconstructed from photons that have never directly interacted with the object. Since its first demonstration with entangled photon pairs \cite{Strekalov_PRL_1995,Pittman_PRA_1995}, ghost imaging has evolved into a general correlation imaging technique that can be implemented not only with quantum optics but also with classical thermal light  \cite{Bennink_PRL_2002,Gatti_PRA_2004,Gatti_PRL_2004,Cai_PRE_2005,Valencia_PRL_2005,Ferri_PRL_2010,Borghi_PRL_2006,Ferri_PRL_2005,Hardy_PRA_2013,Liu_Optica_2021,Sun_Science_2013}, temporal correlations  \cite{Ryczkowski_NP_2016}, and diverse illumination sources ranging from infrared  \cite{Radwell_Optica_2014,Shu_Optica_2026} and X-rays  \cite{Pelliccia_PRL_2016,Yu_PRL_2016} to sunlight  \cite{Liu_OL_2014,Xing_AP_2026} and even to matter waves  \cite{Jeltes_Nature_2007,Khakimov_Nature_2016,Li_PRL_2018,He_SB_2021}. These advances have established ghost imaging as a versatile imaging modality with applications across optics and beyond \cite{Shih_Springer_2012}.

\begin{figure}[htb]
	\centering
	\includegraphics[width=0.81\linewidth]{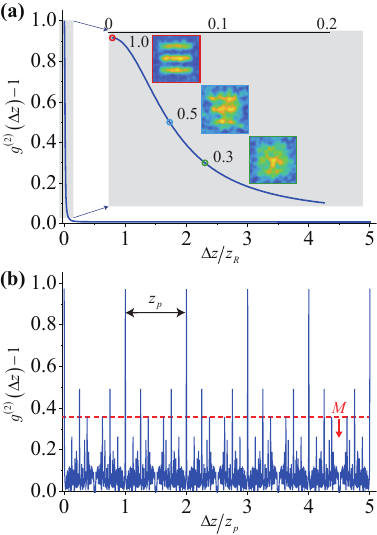}
	\caption{(color online). Longitudinal intensity correlation, $g^{(2)}(\Delta z)-1$, between pairs
    of points located in two transverse planes separated by a longitudinal distance
    $\Delta z$ for (a) conventional thermal incoherent light and
    (b) structured thermal incoherent light. For each pair, one point is located in plane $z=z_1$, while the other in plane $z=z_1+\Delta z$.
    (a) The correlation exhibits a sharp Lorentzian decay,
    $\left[1+(\Delta z/z_c)^2\right]^{-1}$, where
    $z_c=\pi\sigma_c^2/\lambda$ and $\sigma_c$ denotes the transverse
    coherence width of light, as a function of $\Delta z/z_R$.
    The background highlights the corresponding ghost-image recurrences
    at different values of $\Delta z/z_R$.
    (b) The structured coherence-comb exhibits periodic correlation revivals
    at integer values of $\Delta z/z_p$.
    For comparison, the tooth pitch $z_p$ in (b) is chosen to be identical
    to the Rayleigh range $z_R$ in (a).
    The ratio of the transverse coherence width to the beam width in (a)
    is $0.1$, and $M=15$ in (b).}
	\label{fig:1}
\end{figure}

A fundamental limitation, however, has remained essentially unchanged since the inception of thermal ghost imaging: successful image reconstruction requires precise conjugate matching between the reference and object arms, such that the object must remain at a unique in-focus plane \cite{Bennink_PRL_2002,Gatti_PRA_2004,Gatti_PRL_2004,Cai_PRE_2005,Valencia_PRL_2005,Ferri_PRL_2010,Borghi_PRL_2006,Ferri_PRL_2005,Cao_PRA_2005,Scarcelli_APL_2006}. Even a small out-of-focus displacement causes the image to degrade rapidly into a blurred background \cite{Ferri_APL_2008}, severely limiting the flexibility and practical applicability of ghost imaging. This behavior stems from the intrinsic longitudinal coherence [Fig.~\ref{fig:1}(a)] of thermal light in the near-field region, as first demonstrated by Gatti \emph{et al.}~\cite{Gatti_PRA_2008,Magatti_PRA_2009}. Governed by diffraction, the longitudinal intensity correlation follows a rapid Lorentzian decay over a very short out-of-focus distance, as shown in Fig.~\ref{fig:1}(a), thereby confining the imaging capability to a single, narrow object plane. Due to this inherent limitation, ghost image revivals have so far been experimentally demonstrated only for periodic objects in both classical \cite{Luo_PRA_2010,song_PRA_2010} and quantum \cite{Song_PRL_2011,Luo_PRA_2009} ghost imaging.

In this Letter, we lift this fundamental constraint and report the first experimental observation of ghost image revivals for arbitrary---periodic or non-periodic---objects. We achieve this by engineering the longitudinal intensity correlation of structured thermal light fields into a comb-like coherence profile [Fig.~\ref{fig:1}(b)]. Instead of a single correlation peak, the structured thermal light exhibits a periodic sequence of longitudinal correlation peaks, creating multiple discrete in-focus planes where the ghost image faithfully revives.

We begin by representing the instantaneous field of a (pseudo-) thermal light source as a stochastic superposition of coherent modes
\begin{equation}
U(\Br,z)
=
\sum_{\nu} a_{\nu}\Psi_{\nu}(\Br,z),
\label{eq:1}
\end{equation}
where $z$ and $\Br$ denote the longitudinal coordinate and transverse vector
position, respectively. The coefficients $\{a_{\nu}\}$ are mutually
uncorrelated random amplitudes satisfying the second-order statistics
$\langle a_{\nu}a_{\nu'}^{*}\rangle
=\kappa_{\nu}\delta_{\nu\nu'}$, where the angle brackets denote ensemble
averaging, and $\{\kappa_{\nu}\geq 0\}$ represent the modal weights of the
individual modes $\{\Psi_{\nu}\}$. If the spatial modes
$\{\Psi_{\nu}\}$ are chosen as plane waves, Eq.~\eqref{eq:1}
reduces to a conventional thermal light source~\cite{Goodman}, whose
longitudinal coherence exhibits a Lorentzian profile shown in
Fig.~\ref{fig:1}(a). We now specify a particular class of spatial
modes associated with chaotic multimode waveguides or nondiffracting
solutions of the Helmholtz equation in free space as
$\Psi_{\nu}(\Br,z)=\psi_{\nu}(\Br)e^{i\beta_{\nu}z}$, where
$\beta_{\nu}$ is the propagation constant of the mode. We can express the (second-order) mutual coherence function of the structured thermal light fields at pairs of points located in two transverse planes $z=z_{1}$ and $z=z_{1}+\Delta z$ as

\begin{equation}
G(\Br_{1},\Br_{2},\Delta z)
=
\sum_{\nu}
\kappa_{\nu}
\psi_{\nu}(\Br_{1})
\psi_{\nu}^{*}(\Br_{2})
e^{i\beta_{\nu}\Delta z}.
\label{eq:2}
\end{equation}
The analysis of Eq.~\eqref{eq:2} reveals that the correlation in different transverse planes manifests longitudinal dynamics due to the coherence beats of phasors $\exp(i\beta_{\nu}\Delta z)$. In particular, we choose the paraxial Bessel modes $\Psi_{m,\ell}(\Br,z)=J_{\ell}(k_m r)e^{i\ell\phi}e^{-ik_m^{2}z/2k}$ indexed by the integer pair $\{\nu\}=\{m,\ell\}$, where $k_m=2m\pi/d$, with $d$ being a characteristic transverse scale of each mode. The corresponding nonnegative modal weights are set as $\kappa_{m,\ell}=1-(-1)^{\ell}$. It then follows that, for $\Br_1=\Br_2=\Br$, Eq.~\eqref{eq:2} reduces to \cite{Liu_Prapp_2023}
\begin{equation}
G(\Br,\Delta z)
\propto
\sum_{m=1}^{M}
\left[
1-J_{0}\left(\frac{4\pi mr}{d}\right)
\right]
e^{-i2\pi m^{2}\Delta z/z_{p}},
\label{eq:3}
\end{equation}
where $z_p=2d^2/\lambda$, whose physical meaning will be discussed below. Applying the Siegert relation~\cite{Goodman}, we can write for the longitudinal intensity correlation the expression:
\begin{equation}
g^{(2)}(\Br,\Delta z)
=
1+
\frac{\left|G(\Br,\Delta z)\right|^{2}}
{\left|G(\Br,0)\right|^{2}}.
\label{eq:4}
\end{equation}

Figure~\ref{fig:1}(b) illustrates the theoretical distribution of $g^{(2)}(\Br,\Delta z)-1$, calculated from Eq.~\eqref{eq:4}, at the transverse position $r=0.5\,\mathrm{mm}$ for a structured thermal light field composed of $M=15$ modes. In contrast to unstructured thermal light used in conventional ghost imaging [Fig.~\ref{fig:1}(a)], the longitudinal intensity correlation exhibits a series of extremely narrow, periodically recurring peaks separated by the pitch $z_p$, forming a comb-like coherence profile. The fields radiated by this structured thermal light are strongly correlated at each comb tooth, which therefore defines an in-focus plane for ghost imaging. The residual correlation peaks are substantially weaker and can be further suppressed by increasing the number of modes $M$.
\begin{figure}[b]
	\centering
	\includegraphics[width=\linewidth]{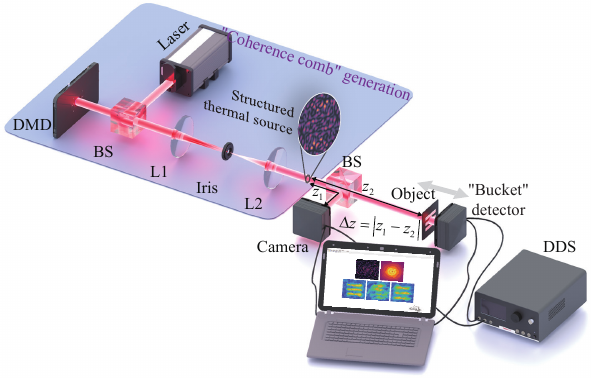}
	\caption{(color online). Schematic of the experimental setup for demonstrating ghost-image
   revivals using a structured thermal light source. DMD, digital micromirror
    device; L1--L2, lenses; BS, beam splitter; DDS, direct digital synthesis
    signal generator. The reference arm is fixed at the length of
    $z_1=5\,\mathrm{cm}$, while the object is translated along the propagation
    direction to a distance $z_2$ from the source. The optical path difference
    between the two arms is $\Delta z=\left|z_1-z_2\right|$.}
	\label{fig:2}
\end{figure}

\begin{figure*}[t]
	\centering
	\includegraphics[width=0.9\linewidth]{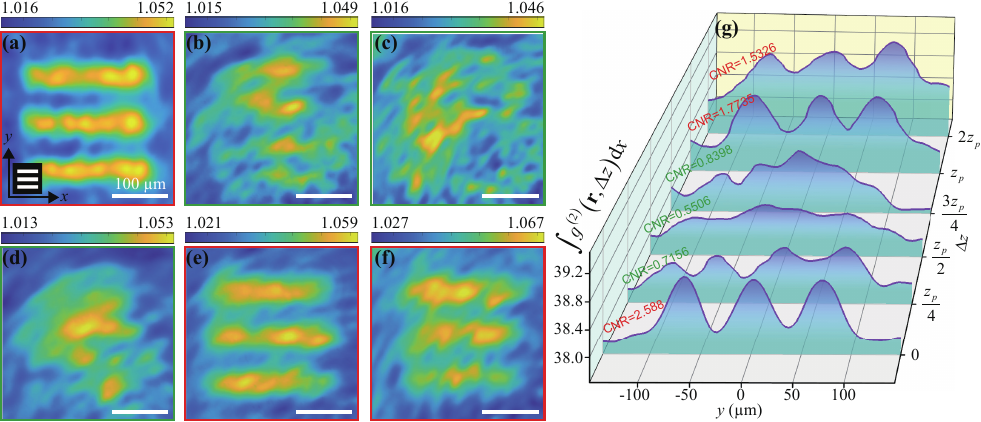}
	\caption{(color online). Experimental results of ghost imaging revivals.
(a)--(f) Measured second-order intensity correlation functions,
$g^{(2)}(\Br,\Delta z)$, obtained from 5000 frames for different
optical path differences:
(a) $\Delta z=0$,
(b) $\Delta z=z_p/4$,
(c) $\Delta z=z_p/2$,
(d) $\Delta z=3z_p/4$,
(e) $\Delta z=z_p$, and
(f) $\Delta z=2z_p$.
(g) Intensity-correlation profiles,
$\int g^{(2)}(\Br,\Delta z)\,\mathrm{d}x$,
of the corresponding images in (a)--(f), demonstrating ghost-image
revivals as the object is translated along the propagation direction
away from the reference plane.}
	\label{fig:3}
\end{figure*}

Next, we validate this theoretical prediction by carrying out a ghost imaging experiment with the structured coherence comb. The experimental setup is sketched in Fig.~\ref{fig:2}. It is based on a standard thermal light ghost imaging geometry, but with two key modifications. (i) The conventional thermal source is replaced by our structured thermal light source generated with a DMD, which emulates a programmable dynamic diffuser. (ii) The radiated incoherent light is then divided by a nonpolarizing BS into a tunable object arm and a fixed reference arm. In the object arm, the beam illuminates a translatable object located at a distance $z_2$ from the source. The object is positioned off axis by 0.2 mm from the optical axis to avoid overlap between the ghost image and its mirror-symmetric “Janus” counterpart (see Supplemental Material~\cite{Suppl}). The transmitted light is then collected by a bucket detector, producing a spatially integrated intensity signal. In the reference arm, the beam propagates over a fixed distance $z_1$, and its spatial intensity distribution is recorded by a high-resolution camera. The optical path mismatch between the two arms is defined as $\Delta z=\left|z_1-z_2\right|$. In the experiment, we used a cw laser operating at $\lambda=532\,\mathrm{nm}$, and the generated structured thermal light field was composed of $M=7$ modes with $d=0.4 \,\mathrm{mm}$, yielding a coherence comb with a pitch of $\sim 60.15\,\mathrm{cm}$. The DMD was refreshed at 250 Hz, and both detectors were synchronized by a DDS at the same frequency as the DMD.

Figure~\ref{fig:3} presents the ghost imaging results obtained with the above experimental setup. We quantify reconstruction quality by the contrast-to-noise ratio (CNR) \cite{Chan_OE_2010}. Whenever $\Delta z=0$, corresponding to an in-focus condition as in early ghost imaging experiments, the intensity autocorrelation reveals a high-fidelity image of the object with CNR=2.5880 [Fig.~\ref{fig:3}(a)]. As the object is longitudinally displaced from the first comb tooth to $\Delta z=z_p⁄4$, the reconstructed image becomes blurred [Fig.~\ref{fig:3}(b)]. The objects in Figs.~\ref{fig:3}(b), \ref{fig:3}(c) and \ref{fig:3}(d) are located between the first and second coherence-comb teeth, where the two arms are out of focus and the object become unrecognizable, with the CNR reduced below 1. As the object reaches the second and third comb teeth, corresponding to $\Delta z=z_p$ and $2z_p$, respectively, the in-focus condition for ghost imaging is restored. As a result, the ghost image reappears with high CNR values (1.7735 and 1.5326), as shown in Figs.~\ref{fig:3}(e) and \ref{fig:3}(f). Figure~\ref{fig:3}(g) shows the transverse profiles of the integrated intensity correlation, $\int g^{(2)}(\Br,\Delta z)\,\mathrm{d}x$, extracted from the reconstructed images in Figs.~\ref{fig:3}(a)--\ref{fig:3}(f). The profiles clearly show ghost image revivals with unit magnification whenever the longitudinal displacement $\Delta z$ equals an integer multiple of the coherence-comb pitch $z_p$.

\begin{figure}[htb]
	\centering
	\includegraphics[width=0.85\linewidth]{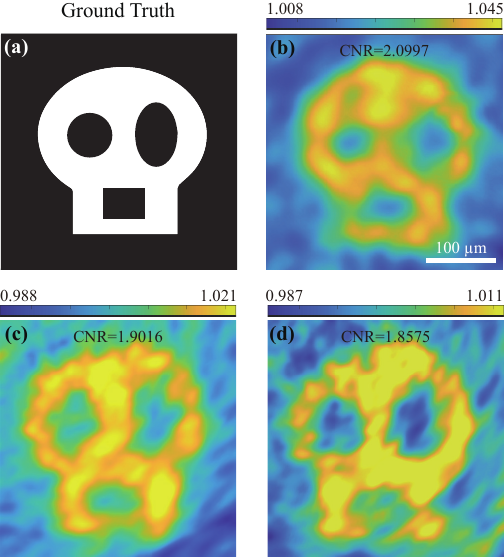}
	\caption{(color online). Experimental observation of ghost imaging revival for a complex
nonperiodic object.
(a) Ground-truth image of the object used in the experiment.
(b)--(d) Reconstructed images obtained with the object plane located at
(b) $\Delta z=0$,
(c) $\Delta z=z_p$, and
(d) $\Delta z=2z_p$.}
	\label{fig:4}
\end{figure}

Finally, we stress that the ghost image revivals observed here are not object-specific. To demonstrate this generality, Fig.~\ref{fig:4} presents our experimental results for a complex, non-periodic ghost-like object [shown in Fig.~\ref{fig:4}(a)]. Clear ghost images [Figs.~\ref{fig:4}(b)--\ref{fig:4}(d).] are recovered at the first, second, and third coherence-comb teeth, corresponding to (b) $\Delta z=0$, (c) $\Delta z=z_p$ and (d) $\Delta z=2z_p$, respectively. In all cases, the reconstructed images exhibit high fidelity, with CNR values exceeding 1.5. In addition, the experimental observation of ghost image revivals for a graded-transmittance object are displayed in the Supplementary Material~\cite{Suppl}.

In conclusion, we have experimentally demonstrated ghost image revivals by tailoring the longitudinal intensity correlation of a structured thermal light field into a coherence comb. Object images are reconstructed whenever the optical path mismatch between the two arms equals an integer multiple of the comb teeth pitch, whereas the images become blurred away from the comb teeth. These results go beyond early ghost imaging experiments based on a single in-focus plane and establish multiplane ghost image revivals for arbitrarily shaped objects. We also emphasize that a quantum version of the ghost image revivals can also be contemplated using entangled photon pairs \cite{Vidal_PRA_2008,Yan_PI_2026}, and that similar revival phenomena can arise in the context of matter-wave physics \cite{Pelliccia_PRL_2016,Yu_PRL_2016,Jeltes_Nature_2007,Khakimov_Nature_2016,Li_PRL_2018,He_SB_2021}. The multiplane ghost image retrieval of arbitrary objects is enabled by the structured thermal source with its distinctive coherence comb profile, which offers the potential for enhanced depth perception in passive imaging systems. This capability may also facilitate ghost imaging-based tomographic reconstruction~\cite{Kingston_Optica_2018}, thereby extending conventional single-plane ghost imaging to the longitudinal dimension. In our experiment, we were able to retrieve high-quality ghost images by translating the object in the range of $-3z_p$ to $3z_p$. The accessible longitudinal imaging range is governed by the pitch of the coherence-comb, but is ultimately limited by diffraction-induced low-pass spatial-frequency filtering, which progressively degrades the image quality at the higher-order coherence-comb teeth. Finally, the introduced concept extends naturally to temporal ghost imaging via engineering the coherence structure of ultrafast laser pulses~\cite{Ryczkowski_NP_2016,Liu_AP_2026,Shih_Arxiv_2025,Joshi_PRA_2024}. In this temporal configuration, propagation through a weakly dispersive medium can further improve both retrieval resolution and range.\\

\emph{Acknowledgments}—Y.C. and X.L. acknowledge funding by the National Natural Science Foundation of China (Grants No. W2441005, No. 12534014, No. 12192254, No. 12547149), the National Key Research and Development Program of China (Grant No. 2022YFA1404800), the Natural Science Foundation of Shandong Province (Grant No. ZR2025ZD21), and by the China Postdoctoral Science Foundation (Grant No. 2026M793716) and the Young Talent of Lifting Engineering for Science and Technology in Shandong (Grant No. SDAST2026 QTA028). S.A.P. acknowledges funding by the Natural Sciences and Engineering Research Council of Canada (Grant No. RGPIN-2025-04064).\\

\emph{Data availability}—The data that support the findings of this paper are not publicly available upon publication because it is not technically feasible and/or the cost of preparing, depositing, and hosting the data would be prohibitive within the terms of this research project. The data are available from the corresponding authors upon request.


\balance

\clearpage
\makeatletter
\renewcommand{\theequation}{S\arabic{equation}}
\renewcommand{\thefigure}{S\arabic{figure}}
\renewcommand{\thetable}{S\arabic{table}}
\pagebreak
\widetext
\begin{center}
\textbf{\large Supplemental Material: Experimental Observation of Ghost Image Revivals via Structured Coherence}\\
\end{center}

\setcounter{equation}{0}
\setcounter{figure}{0}
\setcounter{table}{0}
\setcounter{page}{1}

\noindent {\textbullet \bf Emergence of the “Janus” images in Ghost imaging revivals experiment.}

\begin{figure}[htpb]
	\centering
	\includegraphics[width=\linewidth]{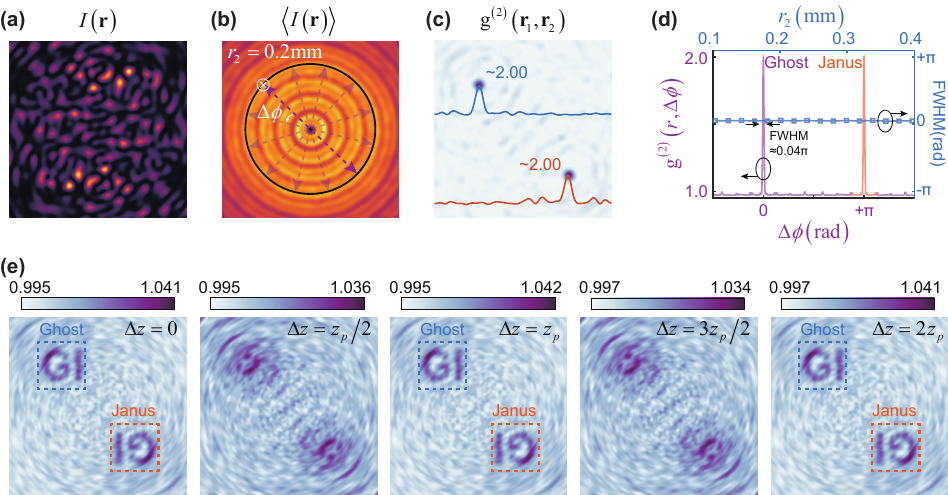}
	\caption{(color online).  Autocorrelation properties of the structured thermal light and the emergence of “Janus” images. (a) Instantaneous speckle-intensity distribution. (b) Ensemble-averaged intensity distribution. (c) Second-order intensity autocorrelation distribution. (d) Angular profile of the second-order autocorrelation function. (e) Simulated ghost image revivals of a “GI” shape object at different longitudinal separations, showing the simultaneous emergence of the real ghost images (blue boxes) and their mirror-symmetric “Janus” counterparts (yellow boxes).}
	\label{fig:S1}
    \end{figure}

In our experiment, we employ the paraxial Bessel modes $\Psi_{m,\ell}(\mathbf{r},z)=J_{\ell}(k_m r)e^{i\ell\phi}e^{-k_m^2 z/2k}$ to synthesis the structured thermal source, Fig.~\ref{fig:S1}(a) shows a representative instantaneous speckled Intensity. Substituting these modes into Eq.~\eqref{eq:1} of the main text and choosing the modal weights as $\kappa_{m,\ell}=1-(-1)^{\ell}$, the generalized field correlation function can be written as
\begin{equation}
G(\mathbf{r}_1,\mathbf{r}_2,\Delta z)
=
\sum_{m=1}^{M}
\kappa_{m,\ell}
J_{\ell}(k_m \left|\bf r_1\right|)
J_{\ell}(k_m \left|\bf r_2\right|)
e^{-i\ell(\phi_1-\phi_2)}
e^{-2\pi i m^2\Delta z/z_p}.
\label{eq:S1}
\end{equation}
where the coherence-comb pitch $z_p$ is defined in the main text. The averaged intensity of such structured thermal light is presented in Fig.~\ref{fig:S1}(b). By further applying the well-known addition theorem for Bessel functions,

\begin{equation}
J_0\!\left(k_m\left|\mathbf{r}_1+\mathbf{r}_2\right|\right)
=
\sum_{\ell=-\infty}^{+\infty}
(-1)^{\ell}
J_{\ell}(k_m \left|\bf r_1\right|)
J_{\ell}(k_m \left|\bf r_2\right|)
e^{-i\ell(\phi_1-\phi_2)},
\label{eq:S2}
\end{equation}
we can sum over the index $\ell$ to cast Eq.~\eqref{eq:S1} into the form
\begin{equation}
G(\mathbf{r}_1,\mathbf{r}_2,\Delta z)
=
\sum_{m=1}^{M}
\left[
J_0\!\left(k_m\left|\mathbf{r}_1-\mathbf{r}_2\right|\right)
+
J_0\!\left(k_m\left|\mathbf{r}_1+\mathbf{r}_2\right|\right)
\right]
e^{-2\pi i m^2\Delta z/z_p}.
\label{eq:S3}
\end{equation}

For $\mathbf{r}_1=\mathbf{r}_2=r e^{i\phi}$,
Eq.~\eqref{eq:S3} reduces directly to Eq.~\eqref{eq:1}
of the main text. We now consider the angular field correlations
between two transverse planes under the condition
$|\mathbf{r}_1|=|\mathbf{r}_2|=r$.
In this case, the field correlation function follows readily from
Eq.~\eqref{eq:S3} as

\begin{equation}
G(r,\Delta\phi,\Delta z)
=
\sum_{m=1}^{M}
\left[
J_{0}\!\left(2k_m r\sin \Delta\phi/2\right)
+
J_{0}\!\left(2k_m r\cos\Delta\phi/2\right)
\right]
e^{-2\pi i m^{2}\Delta z/z_p}.
\label{eq:S4}
\end{equation}
where $\Delta\phi=\phi_1-\phi_2$.\\

Analysis of Eq.~\eqref{eq:S4} reveals that the fields are strongly correlated for $\Delta\phi=0,\pi$ and only weakly
correlated otherwise. Figure~\ref{fig:S1}(c) presents the second-order autocorrelation distribution evaluated at $r_2=0.2\,\mathrm{mm}$, revealing two distinct regions of strong correlation. Equivalently, as shown in Fig.~\ref{fig:S1}(d), the angular autocorrelation function exhibits two narrow peaks (the full-width at half maximum (FWHM) of the peak $\approx 0.04 \pi$): one corresponding to the point itself and the other to its mirror-symmetric counterpart. We refer to the image associated with $\Delta\phi=0$ as the real ghost image and to the mirror-symmetric counterpart as the “Janus” image in our experiment. Figure~\ref{fig:S1}(e) shows the simulated revival of the ghost images at different longitudinal separations $\Delta z$, demonstrating that the “Janus” image appears simultaneously with the real ghost image. Therefore, to prevent spatial overlap between the two images, the target object was positioned off the optical axis in the experiment.

\end{document}